# Energy-dependent photoemission delays from noble metal surfaces by attosecond interferometry

Reto Locher,[1,†] Luca Castiglioni,[2,†,*] Matteo Lucchini,[1] Michael Greif,[2] Lukas Gallmann,[1] Jürg Osterwalder,[2] Matthias Hengsberger,[2] and Ursula Keller[1]

[1]Physics Department, ETH Zurich, CH-8093 Zürich, Switzerland

[2]Department of Physics, University of Zurich, CH-8057 Zürich, Switzerland

[†] These authors contributed equally

*Corresponding author: luca.castiglioni@physik.uzh.ch

**How quanta of energy and charge are transported on both atomic spatial and ultrafast time scales is at the heart of modern technology. Recent progress in ultrafast spectroscopy has allowed us to directly study the dynamical response of an electronic system to interaction with an electromagnetic field. Here, we present energy-dependent photoemission delays from the noble metal surfaces Ag(111) and Au(111). An interferometric technique based on attosecond pulse trains is applied simultaneously in a gas phase and a solid state target to derive surface-specific photoemission delays. Experimental delays on the order of 100 as are in the same time range as those obtained from simulations. The strong variation of measured delays with excitation energy in Ag(111), which cannot be consistently explained invoking solely electron transport or initial state localization as supposed in previous work, indicates that final state effects play a key role in photoemission from solids.**

## 1. Introduction

The dynamical response of the electronic structure of matter to an electromagnetic stimulation, *e. g.* the absorption of a photon is responsible for many physical properties as well as the chemical reactivity. Underlying electronic processes naturally occur on an attosecond (1 as = $10^{-18}$ s) timescale as a result of the characteristic electron velocities and length scales. Photoelectron spectroscopy has been the preeminent tool to study the electronic structure of atoms, molecules and condensed matter in the past 50 years [1-3]. The energetics of the photoemission process has been understood for a long time [4] but its temporal aspect remained largely unexplored due to the lack of experimental tools with the required attosecond time resolution. The interaction of the outgoing electron with the remaining ion creates a slight delay between photon absorption and electron emission. In the case of photoemission from condensed matter additional many-body effects such as dynamical screening and electron-electron scattering as well as transport come into play, which further contribute to the photoemission delay. Such subtle effects determine the lineshape in photoelectron spectra [5] or the lifetime of quasiparticles [6] such as plasmons and excitons, which is of fundamental importance for semiconductors and photovoltaic devices [7].

Recent progress in ultrafast spectroscopy [8] has allowed us to directly study the dynamics of electrons in the time domain. Attosecond energy and angular streaking [9, 10] and reconstruction of attosecond beating by interference of two-photon transitions (RABBITT) [11, 12] are the currently predominant methods to probe ultrafast dynamics on the attosecond time scale. Interaction of the outgoing electron emitted by the attosecond extreme ultraviolet (XUV) pulse with an intense few-cycle infrared (IR) field leads to formation of sidebands (RABBITT) or changes in the electron momentum (streaking). It has been shown in the atomic case that both RABBITT [11, 12] and attosecond streaking [9] deliver the same temporal information about the photoemission process [13]. Whereas streaking was successfully applied to both isolated noble gas atoms and condensed matter systems, RABBITT has been used exclusively in the gas phase until now to study photoionization of atoms [14] and molecules [15]. The formation of sidebands (SB) due to simultaneous absorption of an XUV and an IR photon has been investigated on a platinum surface but no sub-cycle dynamics were observed [16].

Attosecond streaking experiments revealed characteristic relative delays between photoelectrons emitted from different electronic states in both noble gas atoms [14, 17] and in condensed matter systems [18, 19]. Interestingly, different mechanisms were invoked to explain the observed delays. In the atomic case the scattering of the outgoing electron wave packet at the atomic potential leads to a phase shift and thus a delay, as it was first proposed by Wigner in 1955 [20]. In condensed matter the situation is more complex as photoemission involves three steps: excitation, transport to the surface and escape into the vacuum [21]. In the streaking experiments on tungsten [18] and magnesium [19] surfaces it was presumed that streaking of the electron only occurs at the surface or outside the solid and the observed delays were explained in terms of transport from the site of initial excitation to the surface. The measured relative delay of 110 as between *4f* and conduction band electrons in tungsten has been rationalized by various theoretical models. The different emission times were explained in terms of electron transport [22, 23], penetration of the surface barrier [24], different initial state localization [23, 25] and resonant transitions [26]. A Wigner delay in photoemission from solid surfaces has also been discussed as the consequence of an accumulated phase shift of the propagating wave packet [27] as well as the result of inherent phase-shifts associated with final state effects in photoemission [28]. The fact that so many different models were used to reproduce the experimental findings underlines that the dynamics of photoemission in condensed matter are far from being understood and more experimental data are highly needed.

Here, we report a study of energy-dependent photoemission delays from the noble metal surfaces Ag(111) and Au(111). We extended the RABBITT technique to achieve the first observation of sub-cycle dynamics in a condensed matter system using attosecond pulse trains (APT). In previous time-resolved photoemission experiments [14, 17-19] relative delays between two different initial states of the same physical system were examined. In this work the RABBITT technique is simultaneously applied to argon and to a metal surface. The case of Ar being well-understood [13, 14, 29] it is used to calibrate our setup for the temporal characteristics of the XUV pulse train and its timing relative to the IR pulse. This on-the-fly calibration vastly reduces the susceptibility to experimental instabilities and systematic errors. The simultaneous detection allows us to choose a proper reference to gain access to surface-specific photoemission delays without the need for an intrinsic reference state.

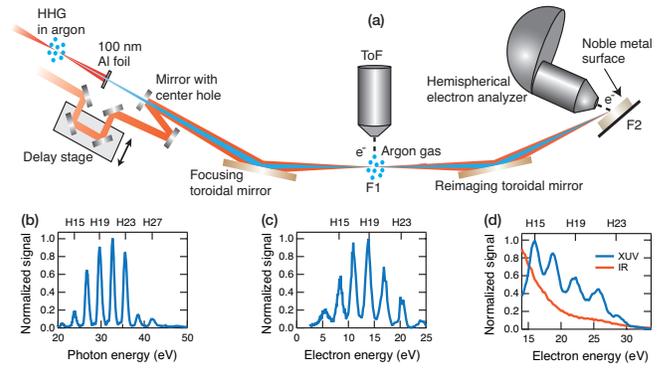

**Fig. 1.** (a) Schematic illustration of the experiment. (b) Typical photon spectrum of the XUV pulse train. (c) Photoelectron spectrum of Ar.
(d) Photoelectron spectra of Ag(111). Replicas of the *4d*-band produced by the harmonics of the XUV pulse sit on a background of secondary electrons (blue line). Moreover, the IR field alone generates an ATP background comparable in strength to the signal of interest (red line).

## 2. Simultaneous RABBITT measurements in two targets

In order to study photoemission dynamics in solid surfaces the existing attosecond beamline was extended with a surface physics endstation comprising a hemispherical electron analyzer [30]. Fig. 1 shows the experimental setup together with typical photon and photoelectron spectra. XUV attosecond pulse trains are produced by high-harmonic generation in argon. Residual IR and low-order harmonic radiation is blocked by a 100 nm Al filter before recombination with the probe beam. A toroidal mirror focuses the co-propagating XUV and IR beams into the source of a time-of-flight (ToF) spectrometer where gas phase RABBITT traces are recorded. A second toroidal mirror images the first focus onto a solid sample surface in the source of a hemispherical electron analyzer where RABBITT traces of the metal surfaces are recorded. Both pulses were *p*-polarized and the angle of incidence on the surface was 75°. Efficient differential pumping kept the pressure in the surface chamber below $7\times10^{-10}$ mbar during the measurements, which allowed us to record RABBITT traces in Ar and on metal surfaces simultaneously. Ag(111) and Au(111) single crystals were cleaned by cycles of sputtering and annealing and the surface quality was verified by XPS and LEED [details in Section 1 of Supplement 1].

RABBITT offers a temporal resolution comparable to attosecond streaking [9] without the need for a single attosecond pulse with its experimental complexity. The lower intensity requirement of the IR probe field leads to reduced perturbation of the system under study. This renders the method less susceptible to above-threshold photoemission (ATP), which enables access to lower photon energies. In contrast to attosecond streaking, photoemission delays are measured for several photon energies (one per sideband) in one single measurement under the same conditions and for the same initial state. The bandwidth of individual harmonics in the APT of around 1 eV was sufficiently narrow to observe reasonable separation between adjacent replicas in the photoelectron spectra of the investigated surfaces.

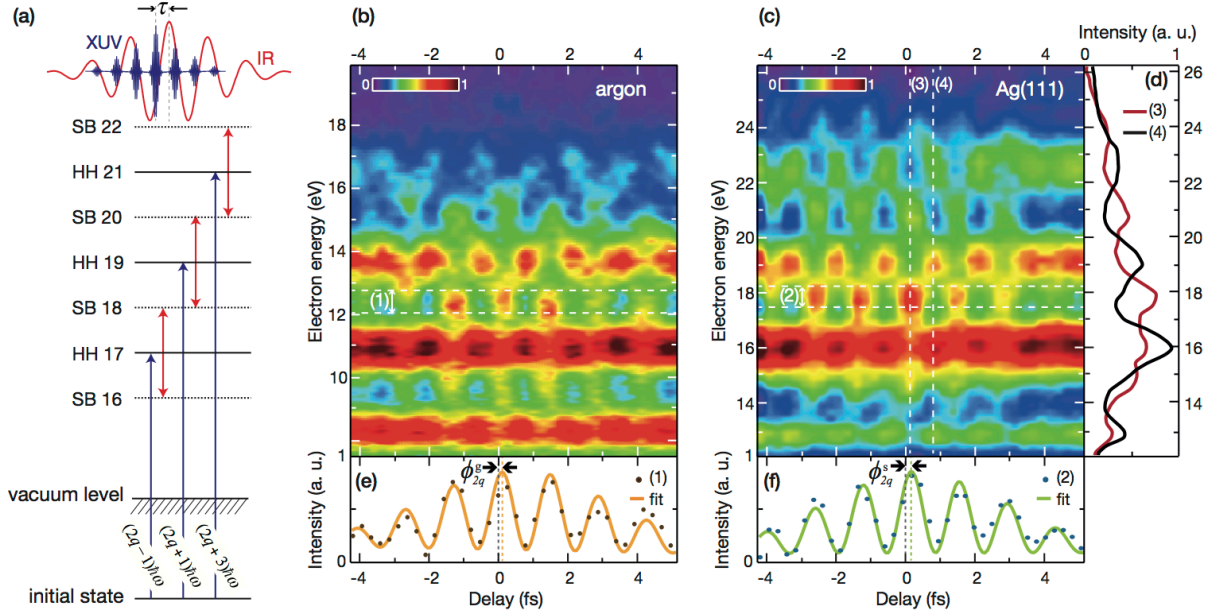

**Fig. 2.** (a) Energy level scheme of the RABBITT process. Interfering two-color two-photon transitions give rise to sidebands (SB) between adjacent odd high harmonics (HH). (b),(c) Experimental RABBITT traces from Ar and Ag(111) with electrons originating from Ar 3p and Ag 4d levels, respectively. Both scans were recorded simultaneously with laser parameters optimized for the surface. A delay-independent background of ATP and secondary electrons was subtracted from (c) to enhance contrast for illustration purposes. (d) Photoelectron spectra from (c) at two different delays. At 100 as (3) the appearance of sidebands is clearly visible whereas at 800 as (4) the photoelectron spectrum qualitatively resembles the spectrum in absence of the IR field. (e),(f) Integration over the energy range of SB 18 revealing the oscillation with $2\omega$. Experimental curves (1) and (2) were fitted with $A(t)\cos(2\omega t - \phi_{2q})$ where $\phi_{2q}$ is the experimental spectral phase as indicated and $A(t)$ is the pulse envelope function.

## 3. Results

### 3.1. Extraction of surface-specific photoemission phase

A set of experimental RABBITT traces in argon and on Ag(111) is shown in Fig. 2 together with an energy level scheme of the process. In Ar the emitted electrons originate from the 3p-state, in Ag(111) the 4d-band is the main contributor. The spectral photoemission phase, $\phi_{2q}^{g/s}$, contains all temporal information and is retrieved by curve fitting [Fig. 2(e)-2(f)]. The same measurements were repeated for Au(111) with emission from the 5d-band. Clear differences between the two noble metals were observed [see Fig. 3]. Signal background due to secondary electrons and ATP is significantly lower in Au(111) owing to the higher work function of this surface compared to Ag(111).

In general, the work function of a surface is significantly lower than the ionization potential of a noble gas atom. Thus the IR probe field leads to electron emission by ATP at substantially lower intensity. The yield of these electrons strongly decreases with kinetic energy, yet energies of up to 35 eV were observed under experimental conditions [see Fig. 1(d)]. In order to reduce their contribution, relatively low probe intensities were employed (a few times $10^{11}\,\text{W/cm}^2$). In addition to ATP, space-charge effects are more severe in a solid-state target due to the high emitter density and obscure the underlying structure of the spectra. Consequently, the flux of the XUV pump pulse was kept low by strongly reducing the intensity of the IR field driving the high-harmonic generation. These specific requirements lead to an unusual intensity regime for RABBITT measurements where the apparent SB amplitude and the depletion of the parent signal appear higher than in previous work [14]. The formalism behind RABBITT to extract photoemission phases from the SB oscillations requires that these sidebands be produced by two-photon transitions (one XUV and one IR photon). Absorption of multiple IR photons can occur at high IR intensities, opening additional quantum paths that may contribute to the oscillating signal and potentially alter the reconstructed photoemission phase [31]. Such higher-order processes would lead to higher frequency contributions of the SB modulation as well as sidebands at harmonic energies exceeding the highest harmonic observed in the XUV spectrum by more than one IR photon. We carefully examined our data but could not find any indication for higher-order processes (see Fig. S1 in Supplement 1). Data sets were taken with varying XUV generation conditions and IR intensities but retrieved phases remained stable, confirming the surface specificity and robustness of our method.

### 3.2. Derivation of photoemission delays

A schematic overview of the RABBITT process for a solid surface is provided in Fig. 4. The Wigner delay, $\tau_\lambda^s$, (1) due to absorption of a XUV photon and transport within the solid, $\tau_{trans}^s$, (2) contribute to the true photoemission delay. The continuum-continuum interaction with the IR probe field (3) yields an additional measurement-induced delay, $\tau_{cc}^s$. The surface-specific photoemission delays, $\tau_{2q}^s$, for Ag(111) and Au(111) shown in Fig. 5 were obtained as follows:

$$\tau_{2q}^s = \tau_{\lambda,2q}^s + \tau_{cc,2q}^s + \tau_{trans,2q}^s$$
$$= \frac{\phi_{2q}^s - \phi_{2q}^g}{2\omega} + \tau_{\lambda,2q}^g + \tau_{cc,2q}^g - \tau_{prop} + \tau_{refl} \quad (1)$$

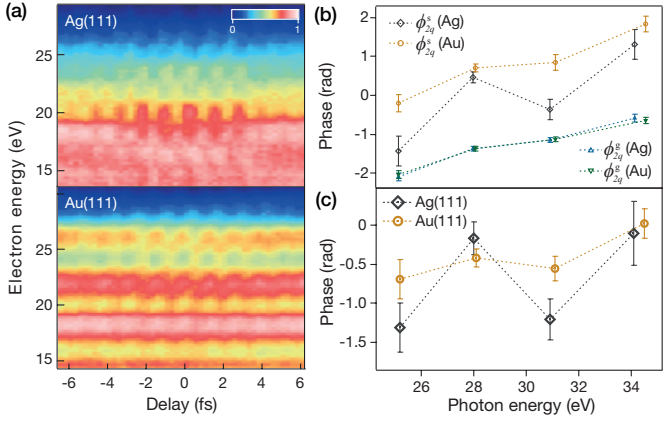

**Fig. 3.** (a) Raw data of typical RABBITT traces from Ag(111) and Au(111). (b) Raw phases, $\phi_2$, extracted from respective surface and argon RABBITT traces. Phases from individual measurements were aligned with the average phase set to zero and contain an unknown offset phase. (c) Surface-specific phase plotted as the phase difference, $\phi_{2q}^s - \phi_{2q}^g$, between corresponding surface and gas phase RABBITT scans. Error bars represent the standard deviation of 6 scans for Ag(111) and 8 scans for Au(111).

$\phi_{2q}^g$ and $\phi_{2q}^s$ are the spectral phases extracted from corresponding surface and gas phase RABBITT traces [see Figs 2(e)-2(f) and 3(b)-3(c)]. $\tau_{\lambda,2q}^g$ and $\tau_{cc,2q}^g$ characterize our temporal reference, the photoemission in Ar, and are taken from literature [13, 29]. The propagation delay between the two targets, $\tau_{prop}$, arises from a phase shift due to reflection at the toroidal mirror and the Gouy phase difference between the first and second focus. This delay was determined experimentally by performing a simultaneous RABBITT measurement with Ar targets in both foci (see Fig. S2 in Supplement 1). The long focal lengths of both toroidal mirrors (1187 mm in F1, 1000 mm in F2) leads to long Rayleigh lengths in F1 and F2 and enables such experimental determination with reasonable precision.

The probe field for the surface measurement consists of the superposition of the incoming IR pulse and the reflected beam, resulting in a transient optical grating. The grazing incidence angle of 15° leads to total reflection and the solid is only penetrated by a weak evanescent field. Conservation of the electric displacement field leads to a sudden drop of the perpendicular component of the electric field at the vacuum-metal boundary due to the high polarization in these nearly free-electron metals. Accordingly, the intensity of the electric field is nearly two orders of magnitude smaller inside the metal right at the boundary [32-34] and the skin depth on the order of 20 nm that describes the exponential decay of the evanescent wave is unimportant regarding the site of absorption of the IR photon. We can conclude that the interaction of the outgoing photoelectron with the IR field must occur right at the surface since both a strong electric field and a steep potential gradient are only present in its close vicinity. The phase of the effective field was calculated based on Fresnel's equations and taking the specific experimental geometry into account, leading to an additional delay, $\tau_{refl}$. An alternative model [35] based on optically determined scattering phases [36] was compared to our model. Reflection phases obtained from the two models agree within 0.2 rad, which corresponds to 43 as in our experiment. The phase of the evanescent wave in a gold surface as a function of incidence angle was determined by photon scanning tunneling microscopy and found to be in good agreement with theory based on Fresnel's equations [37]. Furthermore, X-ray optical effects such as total reflection or standing waves at surfaces and inside solids were recently simulated based on Fresnel's equations and also found to reproduce spectroscopy data very well [38].

While the relative delays between photoemission at different energies can be obtained directly from experimental data, in our case, the calibration of the delay scale relies on two model assumptions, *i.e.* the phase of the effective field at the surface and the description of the temporal reference process [24]. We thus computed the experimental errors for the relative, energy-dependent delays and the absolute time scale separately, since only the zero of the time scale is prone to unknown systematic errors [see Fig. 5(a)-5(b), details in Section 2 of Supplement 1].

### 3.3. Simulation of photoemission delays in solid surfaces

Our theoretical description employs a composite model based on scattering theory and ballistic transport to simulate photoemission in a solid surface. The Wigner delay due to absorption of the XUV photon and the continuum-continuum delay due to interaction with the IR probe field were calculated based on radial matrix elements and corresponding scattering phase shifts that were computed for dipole-allowed transitions in a muffin-tin potential for Ag and Au [39]. Transport times were derived from a ballistic model with group velocities obtained from fitting free-electron final states to the bulk band structure and taking the inelastic mean-free path (IMFP) at given energies into account. The characteristic time scale of screening was predicted to be around 250 as [40]. Hence, we simulated the limiting cases of an unscreened and a completely screened photohole based on a hydrogen-like potential [see Section 3 of Supplement 1].

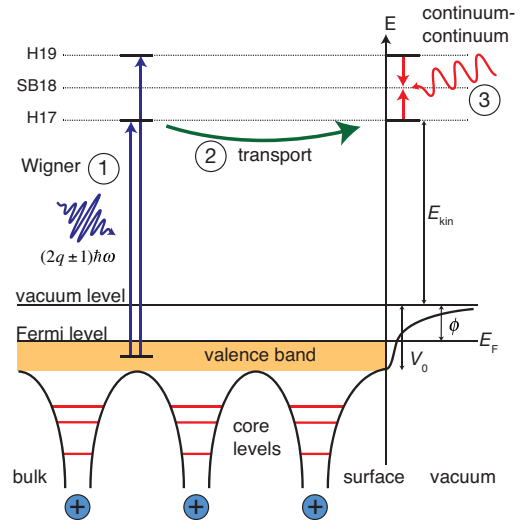

**Fig. 4.** Schematic representation of the three steps involved in the surface RABBITT: (1) initial excitation of the electron by absorption of an XUV photon, (2) ballistic transport within the solid and (3) absorption/emission of an IR photon.

In our case, the difference is small compared to the overall transport delay due to the small effect of the photohole on the deep potential well in Ag and Au. The monotonic decrease of transport times with increasing kinetic energy is a consequence of both the increasing group velocity and the shape of the universal curve of the IMFP in solids [41]. Contributions from the Wigner delay and $\tau_{cc}^s$ are smaller and only slightly alter the shape of the computed energy-delay function.

## 4. Discussion

Calculated and experimental delays in Ag(111) are in good agreement for photon energies of 28 and 34 eV. At other energies, the experimental delays strongly deviate from the model calculations and are much smaller. A possible explanation of this peculiar result could be that XUV excitation in the probed energy region leads to resonant interband transitions to the $\Lambda_6$ $sp$-band [42] close to the $L_{6+}$ van Hove singularity with a high density of states. The availability of bulk final states then leads to enhanced emission from the metal bulk and observed delays are dominated by transport. In the absence of such resonances the photoelectron spectrum is dominated by surface emission [3]. In this regime, electron transport is negligible and observed delays are dominated by $\tau_\lambda^s$ and $\tau_{cc}^s$, resulting in the negative delays observed at 25 and 31 eV. Negative delays were originally predicted by Wigner [20] and also observed in photoemission from Ar atoms [13, 14, 29]. It must be emphasized that photoemission is treated within the framework of scattering theory and can be considered as a half-collision process. Phase shifts are obtained by comparing the scattered outgoing electron wave to a freely propagating wave with same wavevector in the asymptotic limit. Hence the temporal reference is the freely propagating wave and not the absorption of the photon. An absolute delay in terms of time elapsed between absorption of a photon and release of the photoelectron is thus not accessible because the exact position of release is not defined in an infinite-range coulombic potential. We estimated the lower limit for the one-photon delay based on Wigner's causality condition [20, 43] and obtained a value of -70 as for an electron with 30 eV kinetic energy.

Delays in Au(111) show less pronounced excursions from the theoretically predicted, transport-dominated behavior. The $5d$ valence band of Au covers a wider energy range than the $4d$ band in Ag, giving rise to additional interband transitions. Bulk emission is therefore more important in Au at these energies and electron transport cannot be neglected. The interplay between resonant bulk vs. surface emission has been discussed recently [26] to explain the delays observed in earlier streaking experiments [18, 19]. Our experimental data demonstrate that delays in photoemission cannot be solely rationalized by energy-dependent transport times as done in previous studies based on attosecond streaking [18, 19]. Conversely, our results suggest that electron transport is only important if bulk final states are available. Both the energetically broad light sources used in the experiment as well as the width of the initial state complicate the observed dynamics.

It is likely that both resonant and non-resonant transitions contribute to the measured photoemission phase at each harmonic photon energy. Phase shifts and hence different emission times resulting from resonant transitions were also observed in two-photon ionization experiments of molecular nitrogen [15] and helium [44]. Our model employs spherical harmonics final states and is thus unable to reproduce dynamics induced by resonant transitions. Initial state localization[23, 25] has been discussed as a possible origin of different delays observed in the streaking experiments [18, 19]. This can be ruled out in our experiment since we probe photoemission from the same initial state at different photon energies. A more sophisticated theoretical treatment of photoemission in solid surfaces is definitely needed but beyond the scope of this work.

## 5. Conclusions

The ability to sample energy-dependent and surface-specific photoemission delays affords detailed insight into the ultrafast electron dynamics that goes well beyond the measurement of a plain relative delay. Our experimental data demonstrate that neither electron transport nor initial state localization alone can be invoked to rationalize the measured photoemission delays. The strong energy dependence of the delays indicates that photoemission dynamics in this energy range is governed by final state effects. We believe that the RABBITT technique will play a major role in the advancement of attosecond science towards condensed-matter systems as it allows for studying charge dynamics at the inherent electronic timescale with less perturbation from the probe process. The higher energy resolution will allow studying the dynamics of such fundamental processes as spin-orbit interaction in systems with sufficiently large spin-orbit splitting. Furthermore, such experiments access the fastest possible response of an electronic system to interaction with an electromagnetic field and hence provide an upper limit for novel electronic devices in the petahertz regime. and a conservative muffin-tin radius of 2 Å.

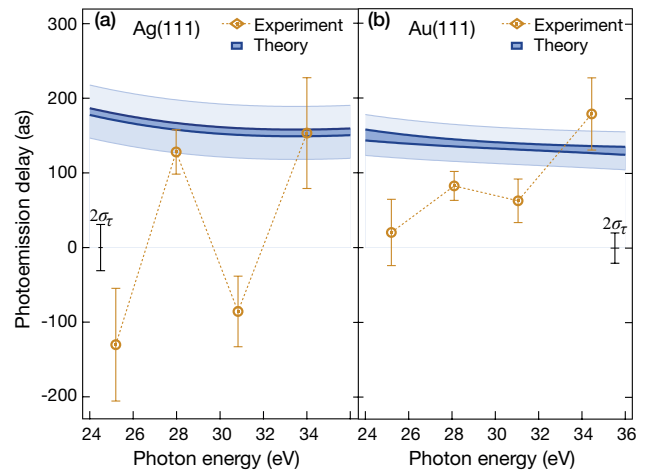

**Fig. 5.** Experimentally determined photoemission delays . (a) Emission from the Ag(111) 4$d$-band (golden diamonds). Results of our simulations are delimited by the extreme cases with and without screening (upper and lower blue line). (b) Results for emission from the Au(111) 5$d$-band. The additional error bars at 0 as in (a) and (b) indicate the experimental error ($2\sigma_\tau$) of the propagation delay, $\tau_{\text{prop}}$, that leads to an uncertainty in the delay scale. This error is added to the simulated values and illustrated by the light-blue shaded areas for better comparison.

## Acknowledgements
We thank P. Krüger for helpful discussions.

## Funding Information
NCCR Molecular Ultrafast Science and Technology (NCCR MUST); Swiss National Science Foundation (SNSF); ETH Zurich Post-doctoral Fellowship Program.

# Energy-dependent photoemission delays from noble metal surfaces by attosecond interferometry

# Supplementary Information

R. Locher,[1]\* L. Castiglioni,[2]\*[†] M. Lucchini,[1] M. Greif,[2] L. Gallmann,[1]
J. Osterwalder,[2] M. Hengsberger,[2] U. Keller[1]

[1]*Physics Department, ETH Zurich, CH-8093 Zürich, Switzerland.*
[2]*Department of Physics, University of Zurich, CH-8057 Zürich, Switzerland.*

\* These authors contributed equally.
[†] Corresponding author: luca.castiglioni@physik.uzh.ch

## 1. Detailed experimental methods

A Ti:sapphire based chirped-pulse amplification system provides ultrashort laser pulses (~25 fs, 1.2 mJ, central wavelength 796 nm) at 1 kHz repetition rate. These pulses are compressed to 7-12 fs by filament compression [1] in two Ar gas cells. The beam is then split into a probe branch (20%) that is sent through a delay line and a pump branch (80%), which produces attosecond pulse trains (20-45 eV, 300 as pulse width) by high-harmonic generation in Ar. After generation of the APT the residual IR and lower-order harmonic radiation is blocked by a 100 nm Al filter. Pump and probe beams are collinearly recombined by means of a drilled mirror; the harmonic radiation passes from behind through the centre hole (diameter 2 mm) of the mirror while the IR probe field is reflected on the front part. A gold-coated toroidal mirror (incidence angle 82°) focuses the two beams in front of a time-of-flight (ToF) spectrometer. Photoemission spectra from Ar gas are recorded at different delays between the two pulses (typical delay step = 250 as). The collection of these spectra constitutes the first RABBITT trace that was used for calibration purpose.

A second toroidal mirror (incidence angle 80°) images the first focus onto a solid sample surface in a one-to-one geometry. Photoemission spectra from the metal surface are recorded with a hemispherical electron analyser. These spectra are measured simultaneously with the ones in Ar and form the second RABBITT trace. Efficient differential pumping kept the pressure in the surface chamber below $7 \times 10^{-10}$ mbar during the measurements, which allowed for prolonged acquisition without degradation of the surface by contamination. Ag(111) and Au(111) single crystals were cleaned by cycles of sputtering and annealing and the surface quality was verified by XPS and LEED. The propagation phase between the two foci was determined experimentally by performing a simultaneous RABBITT experiment with two Ar targets. All measurements were conducted at room temperature and a bias voltage of -5.00 V with respect to the analyzer was applied to the solid sample. The intensity settings for the simultaneous scans were optimized for the surface with a focus on clear SB contrast. Reasonable modulation contrast in the SB signal was obtained with an incidence angle of 75°. All data presented in this work were recorded at this incidence angle. Both pulses are *p*-polarized and the angle of incidence on the surface is 75°.

## 2. Data analysis

### 2.1 Extraction of spectral phase and photoemission delays

In a RABBITT measurement the oscillatory component of the sideband $2q$ is:

$$S_{2q}(t) \sim \cos(2\omega t - \phi_{2q}), \quad (1)$$

where $\omega$ is the angular velocity of the IR driving field, $t$ is the delay between the two pulses and $q$ is a positive integer number. The temporal information about the system and the group delay of the harmonic radiation is encoded in the phase $\phi_{2q}$. In a first step, this phase is retrieved from the experimental RABBITT trace. Subsequently, the contributions from photoemission dynamics and group delay of the harmonic radiation are disentangled.

Our RABBITT scans were recorded with relatively short (7-12 fs) probe pulses and, as a result, the region showing detectable sidebands only covered few laser cycles. A combination of Fourier analysis and curve fitting was applied to retrieve the experimental phase, $\phi_{2q}$, of the sidebands. The signal integrated over the energy range specific to the sideband was Fourier filtered with a Gaussian window of 50 THz width. The peak in the Fourier spectrum of the simultaneously recorded scan in argon was chosen as central frequency. A time-domain filter (super-Gaussian of order 4) and zero padding prior to the Fourier transformation improved the spectral resolution.

For the surface RABBITT scan, instabilities in the XUV flux were removed by subtracting the average surface spectrum weighted by the total counts of the gas spectrum at the given delay. The number of electrons ionized in the gas phase does not depend on the IR field and presents a convenient measure for the flux in the XUV pulse. The filtered (and background subtracted) signal was fitted with the following five-parameter function, both in gas phase and on the surface:

$$S(t) = p_0 e^{p_1(t-p_2)^2} \cos(2\pi p_3 t - p_4). \quad (2)$$

$p_i$ denote the fitting parameters with $p_4$ representing the phase, $\phi_{2q}$, of the corresponding sideband and $p_3$ being twice the central frequency obtained from the Fourier spectrum. It should be noted that the phase, $\phi_{2q}$, could also be retrieved by direct curve fitting without Fourier filtering. In this case a modified fitting function was used that also accounts for a constant background and other contributions to the total signal. The results from both methods agree well within experimental uncertainty.

We performed Fourier analysis of the RABBITT traces to exclude higher-order contributions to the spectral phase. No indication of such higher-order processes can be seen in these spectra (Fig. S1).

Since the sideband oscillation is caused by the interference of two quantum paths, the extracted phase corresponds to the phase difference of these quantum paths and an unknown offset phase. This is written as

$$\phi_{2q}^{g/s} = \theta_{2q+1}^{1/2} + \varphi_{2q+1}^{g/s} + \varphi_{cc,2q+1}^{g/s} - (\theta_{2q-1}^{1/2} + \varphi_{2q-1}^{g/s} + \varphi_{cc,2q-1}^{g/s}) - 2\phi_0^{1/2}. \quad (3)$$

$\theta$ is the phase of the harmonic radiation, $\varphi$ is the system specific dipole transition phase and $\phi_{cc}$ denotes the measurement induced phase caused by the additional IR transition [2, 3]. The subscripts refer to the harmonic order, the superscripts to the target (g=gas, s=surface) or the site (1st and 2nd focus) of the measurement. The offset phase $\phi_0^{1/2}$ relates to the choice of $\tau = 0$, which is not experimentally accessible with required accuracy. Rearranging terms and omitting superscripts in the previous equation yields

$$\begin{aligned}\phi_{2q} &= (\theta_{2q+1} - \theta_{2q-1}) + (\varphi_{2q+1} - \varphi_{2q-1}) + (\varphi_{cc,2q+1} - \varphi_{cc,2q-1}) - 2\phi_0 \\ &= \Delta\theta_{2q} + \Delta\varphi_{2q} + \Delta\varphi_{cc,2q} - 2\phi_0 \\ &= 2\omega \cdot (\tau_{GD,2q} + \tau_\lambda + \tau_{cc} - \tau_0).\end{aligned} \quad (4)$$

$\tau_{GD}$ is the group delay of the high harmonic radiation, $\tau_\lambda$ (single-photon) photoemission delay, $\tau_{cc}$ the continuum-continuum delay caused by the probe field [3] and $\tau_0$ the time-shift corresponding to the offset phase. Since the sideband phase measured in RABBITT corresponds to a finite-difference approximation of the group delay at the sideband energy, the link between the phases and attributed delays becomes apparent. $2\omega$ is the frequency separation of two neighboring sidebands. In the case of the calibration measurement in argon, the quantities $\tau_\lambda$ and $\tau_{cc}$ in Eq. (4) are known from literature [3,4].

### 2.2 Propagation and reflection phase

Care has to be taken when applying this calibration to the scan on the surface. The two measurements are conducted at different sites and the two pulses acquire additional phases during propagation between the two foci of the apparatus. This situation is indicated by the superscripts 1 and 2 in Eq. (3). The difference of the experimental phases in focus 1 and focus 2 can be written as follows:

$$\phi_{2q}^{(2)} - \phi_{2q}^{(1)} = \Delta\theta_{2q} + \Delta\theta_{2q}^{toro} + \Delta\varphi_{2q}^{(2)} + \Delta\varphi_{cc,2q}^{(2)} \\ -2\phi_0^{(2)} - (\Delta\theta_{2q} + \Delta\varphi_{2q}^{(1)} + \Delta\varphi_{cc,2q}^{(1)} - 2\phi_0^{(1)}).$$  (5)

Since both IR and XUV beams propagate in vacuum their phase is only affected by the reflection on the toroidal mirror as well as the Gouy phase in the focus. $\Delta\theta_{2q}^{toro}$ accounts for the propagation phase of the XUV pulse. The offset phase, $\phi_0^{(2)}$, comprises all phases accumulated by the IR pulse between 1st and 2nd focus and is given by

$$\phi_0^{(2)} = \phi_0^{(1)} + \phi_{prop}^{IR} + \phi_{refl}^{IR}.$$  (6)

Noble metal surfaces are highly reflective in the IR and the probe field constitutes a transient grating[5] formed by the superposition of the incoming and reflected IR pulses with an associated phase $\phi_{refl}^{IR}$.

Whereas the reflection phase of the toroidal mirror can be calculated using Fresnel's equations it is rather delicate to compute the Gouy phase. Hence we determined the propagation phase between the two foci experimentally. For this purpose a gas target was installed in the surface chamber at the position where the solid sample was placed normally. This enabled us to simultaneously record RABBITT traces of Ar in the 1st and 2nd focus. Invoking Eq. (5) we see that if we take the difference of the acquired phases we are left only with $\phi_{2q}^{(2)} - \phi_{2q}^{(1)} = 2\phi_0^{(1)} + \Delta\theta_{2q}^{toro} - 2\phi_0^{(2)}$. With a gas target in focus 2 there is no reflection on the surface and thus $\phi_{refl}^{IR} = 0$. We can therefore derive an experimental propagation phase:

$$\phi_{prop,2q} = \phi_{2q}^{(2)} - \phi_{2q}^{(1)} = \Delta\theta_{2q}^{toro} - 2\phi_{prop}^{IR}.$$  (7)

Any remaining energy dependence of the propagation phase must be attributed to the reflection of the XUV on the toroidal mirror since the intrinsic harmonic phases cancel out each other. Delays due to the reflection phase of the XUV in the region of 25-35 eV were computed to be more than two orders of magnitudes smaller than our experimental delays. This allowed us to use an energy-independent mean propagation phase, $\phi_{prop}$, which is indicated by the blue line in Fig. S2.

In order to establish the relation between the offset phases $2\phi_0^{(1)}$ and $2\phi_0^{(2)}$ for surface RABBITT measurements we must assess $\phi_{refl}^{IR}$. We compute the phase of the transient grating due to reflection on the surface based on Fresnel's equations. For parallel polarization the complex reflection coefficient for photon energy $\varepsilon$ and angle of incidence $\theta$ is computed as

$$r_p(\varepsilon) = \frac{\tilde{n}(\varepsilon)^2 \cos\theta - \sqrt{\tilde{n}(\varepsilon)^2 - \sin^2\theta}}{\tilde{n}(\varepsilon)^2 \cos\theta + \sqrt{\tilde{n}(\varepsilon)^2 - \sin^2\theta}},$$  (8)

with complex refractive index $\tilde{n}$. Material data for $\tilde{n}$ were taken from Palik[6]. The Fresnel reflection phase of the IR, $\phi_{Fres}^{IR}$, is directly given by the phase of expression (8):

$$\phi_{Fres}^{IR} = \arg(r_p(\varepsilon)) \tag{9}$$

The phase modulation introduced by the 2D-transient grating was computed as a function of the incident angle $\theta$ of the IR beam and the final momentum of the photoelectrons. Our simulations show that electrons with different directions are affected differently by the transient grating. The effective phase of the IR field felt by electrons with final direction in the detection cone of the analyzer (between 25° and 35° with respect to the surface normal) can be written as

$$\phi_{refl}^{IR} = \phi_{Fres}^{IR}/\beta, \tag{10}$$

where the parameter $\beta = 1.70$ was extracted from simulations taking the specific experimental geometry into account. The simulated transient grating is illustrated in Fig. S3.

### 2.3 Calibrated photoemission delays

All propagation-related phases being assessed, the relationship between corresponding quantities at the different measurement sites can be established. The difference of the experimental phases in focus 1 (gas) and focus 2 (surface) can be rewritten as follows:

$$\phi_{2q}^s - \phi_{2q}^g = \Delta\varphi_{2q}^s + \Delta\varphi_{cc,2q}^s - \Delta\varphi_{2q}^g - \Delta\varphi_{cc,2q}^g + \phi_{prop} - 2\phi_{refl}^{IR} \tag{11}$$

Finally the (two-photon) photoemission delay for the metal surface is obtained by taking the difference, $\phi_{2q}^s - \phi_{2q}^g$, making the link between phases and attributed delays according to Eq. (4) and rearranging the terms:

$$\begin{aligned}\tau_{2q}^s &= \tau_{\lambda,2q}^s + \tau_{cc,2q}^s + \tau_{trans}^s \\ &= \frac{\phi_{2q}^s - \phi_{2q}^g}{2\omega} + \tau_{\lambda,2q}^g + \tau_{cc,2q}^g - \tau_{prop} + \tau_{refl}.\end{aligned} \tag{12}$$

The resulting delay, $\tau_{2q}^s$, is the sum of all processes that contribute to the photoemission from a metal in the RABBITT process, namely a Wigner-type one-photon delay, $\tau_\lambda^s$, a contribution from transport, $\tau_{trans}$, as well as an additional delay, $\tau_{cc}^s$, resulting from the continuum-continuum interaction with the IR probe field.

## 3. Simulation of photoemission delays

### 3.1 Calculation of Wigner delays and $\tau_{cc}$

The Wigner delay arises from the initial excitation by the XUV pulse out of *4d* and *5d* bands in Ag and Au, respectively. This delay originates from the phase of the outgoing intermediate state $|\vec{k}\rangle$ as compared to a plane wave of same wavevector and can be calculated as scattering phase of half an elastic scattering event since photoemission can be considered a half collision. Phase shifts and radial matrix elements were calculated using the Linearized Muffin-Tin Orbital (LMTO) method [7]. For the excitation, dipole selection rules are assumed, *i.e.* excitation into outgoing spherical harmonics of *l*=1 and *l*=3. The coherent superposition of *p*- and *f*-states yields the phase of state $|\vec{k}\rangle$.

The amplitudes of the transition into *p*- or *f*-states, respectively, are calculated by evaluating the transition matrix elements of initial and final states. The time delay associated with the phase is obtained by

$$\tau_{Wigner} = \hbar\frac{d\varphi}{dE}. \tag{13}$$

The interaction of the intermediate state with the IR field described by the transition $|\vec{k}\rangle \to |\vec{k}'\rangle$ leads to $\tau_{cc}$. Similar to the computation of the Wigner delay described above, the sum of the two half-scattering phases yields the phase shift of a partial wave

due to the continuum-continuum transition. The incoming wave (from photoemission from $d$ states) is of $p$- or $f$-character with the amplitudes given by the initial excitation, *i. e.* the radial matrix elements calculated within the LMTO scheme [7]. All contributions were summed up coherently as different magnetic quantum numbers are not distinguished in the experiment.

The total continuum-continuum contribution to the delay is then given by

$$\tau_{cc} = \frac{\varphi_> - \varphi_<}{2\omega}.$$
(14)

Here $\phi_>$ and $\phi_<$ are the phases along two different quantum paths (absorption and stimulated emission) that lead to the same sideband and are obtained by recurring the following formula for all partial wave phase shifts:

$$\tan(\varphi'_<) = \frac{A_{ps} \cdot \sin(\delta_{p \to s}) + A_{pd} \cdot \sin(\delta_{p \to d})}{A_{ps} \cdot \cos(\delta_{p \to s}) + A_{pd} \cdot \cos(\delta_{p \to d})},$$
(15)

with

$$\delta_{p \to s}(E_{SB}) = \delta_p(E_{SB} + \hbar\omega) + \delta_s(E_{SB}).$$
(16)

$\delta_{i \to j}$ denote respective photoemission phase shifts. $\varphi'$ in Eq. (15) is the phase of the coherent superposition of partial waves $s$ and $d$. To coherently superimpose this wave with the next allowed partial wave following path $f$ to $d$ Eq. (15) is recurred. The recursion is repeated until all partial waves are summed up to yield one wave with phase $\phi_<$.

### 3.2 Calculation of transport times

Transport times were modeled based on ballistic transport in the intermediate state after the XUV excitation. The transition momenta (assuming direct transitions due to negligible photon momenta) are obtained from the momentum of the photoelectron in vacuum. Using the formulas below for the refraction at the surface, the momentum in the crystal can be calculated:

$$k_\| = \sqrt{\frac{2m^*}{\hbar^2}(\hbar\omega_{XUV} - E_B - \phi)} \cdot \sin(\Theta)$$
(17)

$$k_\perp = \sqrt{\frac{2m^*}{\hbar^2}\left[(\hbar\omega_{XUV} - E_B - \phi) \cdot \cos^2(\Theta) - V_0\right]}$$
(18)

$V_0$ is the inner potential with respect to the vacuum level, $\Theta$ the emission angle with respect to the surface normal (111), $m^*$ the effective mass of the final state band, and the work function of the samples.

|  | Ag(111) | Au(111) |
|---|---|---|
| $m^*$ | $m_e$ | $m_e$ |
| $V_0$ (eV) | -12.24 | -14.4 |
| (eV) | 4.74 | 5.4 |
| $V_0$ (eV) | -5 | -4 |
| $\Theta$ (deg) | 30 | 30 |

**Table S1**. Values for Eqs (17) and (18) taken from Refs [8, 9].

The group velocity is obtained from fits of free-electron final states to the bulk band structure using the inner potentials and effective masses given in Tab. S1. The group velocity obtained is given by the derivative:

$$v_g = \hbar^{-1}\frac{\partial E(k)}{\partial k} \approx 30 \text{ Å/fs} \tag{19}$$

The (screened or unscreened) hydrogen-like potential of the photohole was included such that the wave vector and group velocities were evaluated numerically as a function of distance from the site of the XUV excitation. The potential was of the form

$$V_{PH}(r) = \frac{-e^2}{4\pi\varepsilon_0 r} \cdot e^{-r/l_{TF}} \tag{20}$$

with the Thomas-Fermi length, $l_{TF}$, of 0.5 Å for silver and 0.2 Å for gold [10]. In the case of the unscreened potential the exponential factor was set to 1.

Integration over the path length yields an effective transport time. The typical time scale of the build-up of screening is predicted to be in the range of 250 as [11], and hence in the range of our experiments. Since the actual time scale is unknown the transport was evaluated once for a completely screened and once for an unscreened photohole potential.

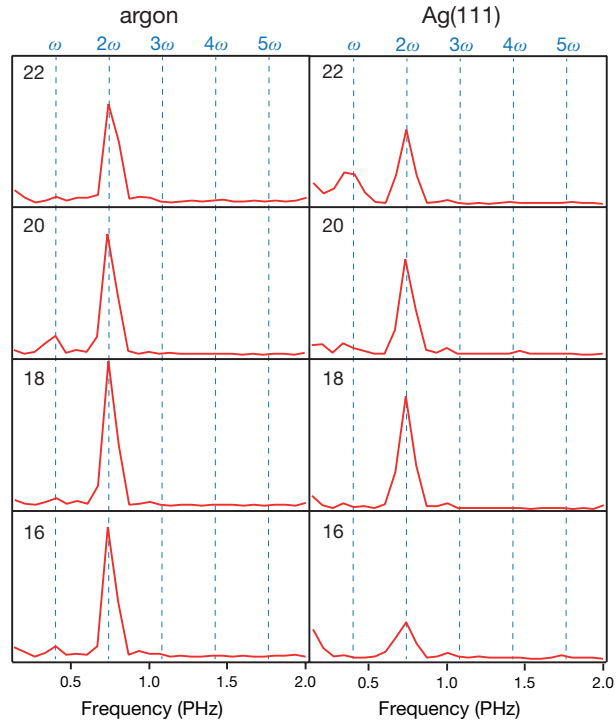

**Fig. S1.** Normalised power spectra of the Fourier analysis of the sideband modulation in argon and Ag(111). The number of the harmonics and frequencies as integer multiples of the fundamental IR frequency are indicated. The $2\omega$ signal is the modulation of the RABBITT process.

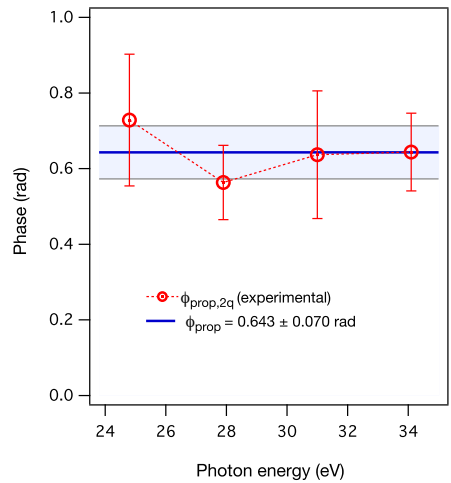

**Fig S2.** Propagation phase derived from RABBITT measurements in two Ar targets at the two different foci. The blue shaded area indicates the standard deviation of the mean phase.

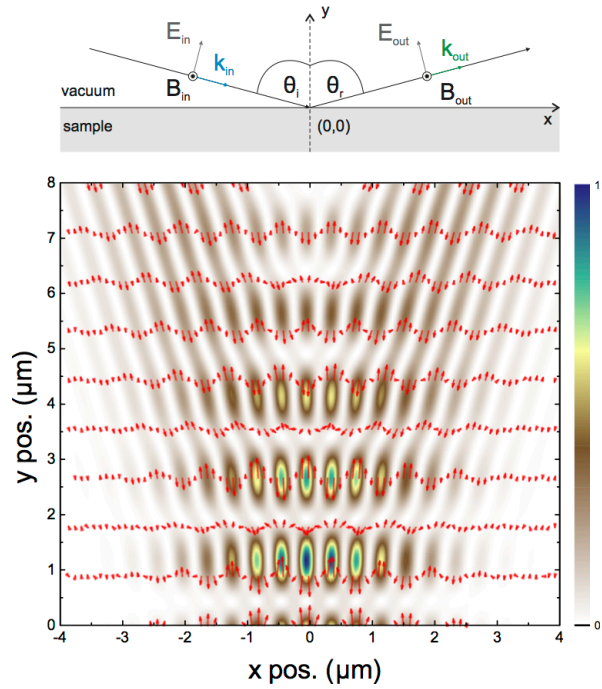

**Fig. S3.** Two-dimensional transient grating due to reflection of the IR pulse at the sample surface at instant $\tau = 0$ fs. The colour scale on the right displays the intensity of the electric field, the amplitude is indicated by the red arrows. The top scheme illustrates the vectors describing the IR pulse and the sign convention used in our calculations.